\documentstyle[12pt]{article}
\begin{document}
\baselineskip=24pt  

\title{Solar Neutrino Decay}
\author{Andy Acker and Sandip Pakvasa
        \\Department of Physics and Astronomy\\
        University of Hawaii at Manoa\\}

\date{}

\maketitle

\begin{abstract}
We re-examine the neutrino decay solution to the solar neutrino problem
in light of the new data from Gallex II and Kamiokande III.
We compare the experimental
data  with the solar models of Bahcall and Pinsonneault and Turck-Chieze
and find that neutrino decay is ruled out as a solution to the solar
neutrino problem at better than the 98\% c.l. even when
solar
model uncertainties are taken into account.
\end{abstract}

\eject

\section{Introduction}

It has been pointed out that
the inflight
decay of solar neutrinos provides a possible
solution to the solar neutrino problem\cite{decay}. Such scenarios have gained
interest as they have been shown capable of providing
simultaneous solutions to both the low
energy atmospheric neutrino anomaly (via neutrino oscillations)
and the solar neutrino problem
(via in flight decay) while requiring mixing only
between 2 generations\cite{simult}. Further, it has recently been shown
that certain models implementing these ideas can lead to an
observable rate of
neutrinoless double beta decay accompanied by majoron emission\cite{cliff}.
In this note we review the phenomenology of solar neutrino decay
and  re-evaluate this solution to the solar
neutrino problem in light of the new results from Gallex\cite{gII} as well as
Kamiokande\cite{kIII}.

\subsection{Neutrino Decay Phenomenology}

We begin by reviewing the phenomenology of solar
neutrino decay. Let us
take $\nu_e$ to be a mixture of mass eigenstates $\nu_i$ with masses $m_i$;
$\nu_e=\sum_i U_{ei}\nu_i$ and assume that one of these, say $\nu_2$, is
unstable with a rest frame lifetime of $\tau_0$.
It is implicit that all other neutrino mass eigenstates have lifetimes
much greater than the Earth-Sun transit time even for the highest solar
neutrino energies.
In the presence of neutrino decay the solar $\nu_e$ flux is
depleted and the spectrum distorted as given by
\begin{equation}
\phi(\nu_e, E)=\phi_{\odot}(E) \times \{(1-|U_{e2}|^2)^2 + |U_{e2}|^4
exp[-t/\tau(E)]\},
\end{equation}
where $\tau(E)$ is the lifetime at energy $E$ $(\tau(E)=(E/m_2)\tau_0)$, $t$
is the Sun-Earth time of flight, about 480 s, and $\phi_{\odot}$ is the
SSM $\nu_e$ flux.
There is, in addition, a $\nu_{\mu}$ flux resulting from $\nu_e$ conversion
which must be accounted for when considering modification to the solar neutrino
signal as measured in electron scattering or neutral-current detectors.
This is given by
\begin{equation}
\phi(\nu_{\mu}, E)=\phi_{\odot}(E) |U_{e2}|^2 \{(1- |U_{e2}|^2)
[1+exp(-t/\tau(E))]\}
\end{equation}
in the limit of two-flavor mixing. Hence the spectral suppression
is completely determined by two parameters, the lifetime $\tau$
and the mixing angle $|U_{e2}|$.

Two models have recently been investigated that give rise to
fast neutrino decay in vacuum
as required by the solar neutrino problem and are consistent with
all laboratory constraints. One class of these models\cite{ddirac}
assumes that
the neutrinos are Dirac particles, and that the coupling which gives rise
to neutrino decay is of the form
$g_{21}\nu^T_{R1}C^{-1}\nu_{R2}\chi$ where $\chi$ is a light iso-singlet
scalar.
This coupling leads to the decay in-flight of
$\nu_2$: $\nu_2 \rightarrow \bar{\nu}_{1R} + \chi$. As $\bar{\nu}_{1R}$ is a
right-handed singlet the decay products in this model are sterile.
A second class of models\cite{dmaj} assumes that neutrinos are Majorana
particles and that the coupling
responsible for neutrino decay is of the form
$g_{21}\nu^T_{L1}C^{-1}\nu_{L2}J$ where $J$ is a Majoron. This coupling
 leads to the in-flight decay of
$\nu_2$: $\nu_2 \rightarrow \bar{\nu}_{1} + J$.
In this case $\bar{\nu}_{1}$ is a superposition of ordinary anti-neutrinos
and interacts as a $\bar{\nu}_e$ with probability $|U_{e1}|^2$.
Hence in this model the initial solar $\nu_e$ flux gives rise to a decay
modulated $\bar{\nu}_e$ flux, as an additional signal for the
decay\cite{neflux}.

We note that a new class of models\cite{mid} has recently been proposed
wherein matter effects can induce
neutrino decay even if the neutrino is stable in vacuum. This can
lead to a very different energy dependence of the solar neutrino flux
suppression\cite{br} than that discussed above in the vacuum decay scenario.
Detailed numerical calculations of the resultant solar fluxes in such models
have not yet been carried out, and we do not comment
on the viability of such models as a solution to the solar neutrino
problem.

\section{Neutrino Decay and Solar Neutrino Data}

We now evaluate the viability of the vacuum neutrino decay solution in light
of the new data from the $^{71}Ga$ experiment Gallex and from Kamiokande.
The combined results of the $^{71}Ga$ experiments, SAGE\cite{sage},
GALLEX I\cite{gI}
and GALLEX II\cite{gII} give $77\pm 13$ SNU's,
the Homestake $^{37}Cl$ experiment
reports $2.28 \pm .028$ SNU's\cite{cl}, and the Kamioka water Cerenkov detector
reports a flux of $ 0.51 \pm .07$\cite{kIII} of the Bahcall
Pinsonneault\cite{bp}
SSM predictions. These experimental results are given in
table 1 as a fraction of predictions of Bahcall Pinsonneault and
Turck-Chieze\cite{tc} SSM's. The error bars in the data are the 1$\sigma$
experimental errors divided by the SSM prediction.
To take the model uncertainties into account,
we also compare the data to the SSMs with the neutrino fluxes at their
$1\sigma$ upper and lower limits. Thus, for example, the
row in table 1 labeled "BP $-1\sigma$" gives the experimental results
as a fraction of the Bahcall Pinsonneault SSM where all neutrino
fluxes are taken to be at the model's $1\sigma$ lower limit.
In addition, we have included
 a comparison of the data to the BP model where
the $^8B$  neutrino flux is at its $2\sigma $ lower limit.

Table 2 shows the best fit parameters of the neutrino decay solution
$|U_{e2}|$ and $\tau$ and the corresponding value of $\chi^2$ for
each of the Solar Models under consideration. Note that in each case
a short lifetime is preferred, we find that, in general, a lab frame
lifetime for a 10 MeV neutrino of 30 seconds or less gives approximately the
same
$\chi^2$.
The best fit occurred
for the SSM of Turck-Chieze with all $\nu_e$ fluxes at their $1\sigma$
lower limit. The minimum $\chi^2$ is 10.7 for three degrees of freedom,
hence this solution is excluded at the 98\% confidence level.
For all other cases tested,
in particular the B.P. and T-C SSM's with all neutrino fluxes at their
central values, we find that the neutrino decay solution to the
solar neutrino problem is ruled out at better than the 99\%
confidence level.

\section{Conclusions}
We have re-examined the neutrino decay solution to the
solar neutrino problem in light of the new data from
Gallex II. We find that the results from the $^{71}Ga$,
$^{31}Cl$, and Kamiokande water Cerenkov detectors can not be
simultaneously explained by the in flight decay of solar
neutrinos. Assuming either the SSM of Bahcall and Pinsonneault
or Turck-Chieze, and taking uncertainties in the
predicted solar neutrino flux into account the decay scenario
is ruled out at greater than the 98\% confidence level.

\begin{center}\subsection*{Acknowledgements}\end{center}
 This work
was supported in part by the U.S. Department of Energy under contract
DE-AM03-76SF00235.

\eject
\section*{Tables}


\begin{table}[bth]\centering
\caption{Data Compared to SSM's}
\label{t1}
\vspace*{0.2in}
\begin{tabular}{|l||c|c|c|}
\hline
Solar Model& Cl&   Kamioka  & Gallium \\
\hline
\hline
BP cental value & .28 $\pm$ .03 &  .51 $\pm$ .07     & .59 $\pm$ .1   \\
\hline
BP $+1\sigma$&  .25 $\pm$ .03 &  .45 $\pm$ .06 & .56 $\pm$ .1 \\
\hline
BP $-1\sigma$&   .33 $\pm$ .03 &  .59 $\pm$ .08 & .61 $\pm$ .1\\
\hline
BP $-2\sigma\; ^8B$&   .37 $\pm$ .04 &  .72 $\pm$ .10 & .60 $\pm$ .1\\
\hline
TC central value&   .36 $\pm$ .04 &  .66 $\pm$ .08 & .61 $\pm$ .1\\
\hline
TC $+1\sigma$&   .29 $\pm$ .03 &  .53 $\pm$ .05 & .56 $\pm$ .1\\
\hline
TC $-1\sigma$&   .46 $\pm$ .05 &  .88 $\pm$ .12 & .66 $\pm$ .1\\
\hline
\end{tabular}
\end{table}

\begin{table}[bth]\centering
\caption{Decay Solution Fits}
\label{t2}
\vspace*{0.2in}
\begin{tabular}{|l||c|c|c|}
\hline
Solar Model& Lifetime& $|U_{e2}|$& $\chi^2$\\
\hline
\hline
BP cental value & 0 &  .635& 13.7\\
\hline
BP $+1\sigma$&  0 & .672& 13.2\\
\hline
BP $-1\sigma$&   0 &.622&12.8\\
\hline
BP $-2\sigma\; ^8B$&0& .583& 11.7\\
\hline
TC central value& 0& .582 &12.3\\
\hline
TC $+1\sigma$&0& .634 & 16.6\\
\hline
TC $-1\sigma$&0 &.514& 10.7\\
\hline
\end{tabular}
\end{table}

\eject

\end{document}